\documentclass[a4paper,11pt]{article}
\usepackage{aaskaiid}
\usepackage{rotating}
\usepackage{orcidlink} 
\usepackage{wrapfig}

\newcommand{\Vr}{\vec{r}}
\newcommand{\ud}{{\rm d}}

\newcommand{\bperp}{B_\perp}

\title{Early phases of star formation with SKAO: synchrotron emission from dense starless cores in molecular clouds}
\ShortTitle{Synchrotron emission from dense starless cores with SKAO}

\author[1,2,3]{Andrea Bracco\,\orcidlink{0000-0003-0932-3140}}
\ShortName{Andrea Bracco et al.} 
\author[4]{Tyler L. Bourke\orcidlink{0000-0001-7491-0048}}
\author[5,6,7,2]{Stefano Bovino}
\author[8]{Clive Dickinson\orcidlink{0000-0002-0045-442X}}
\author[2]{Daniele Galli\orcidlink{0000-0001-7706-6049}}
\author[9]{Michael Kuffmeier}
\author[2]{Marco Padovani\orcidlink{0000-0003-2303-0096}}
\author[10]{Stefania Pezzuto\orcidlink{0000-0001-7852-1971}}
\author[11]{Thushara G.S. Pillai\orcidlink{0000-0003-2133-4862}}
\author[2]{Giovanni Sabatini\,\orcidlink{0000-0002-6428-9806}}
\author[10]{Alessio Traficante}

\affiliation[1]{LUX, Observatoire de Paris, Université PSL, Sorbonne Université, CNRS, 75014 Paris, France}
\affiliation[2]{INAF – Osservatorio Astrofisico di Arcetri, Largo E. Fermi 5, 50125 Firenze, Italy}
\affiliation[3]{Laboratoire de Physique de l'Ecole Normale Sup\'erieure, ENS, Universit\'e PSL, CNRS, Sorbonne Universit\'e, Universit\'e de Paris, F-75005 Paris, France}
\affiliation[4]{SKA Observatory, Jodrell Bank, Lower Withington, Macclesfield, Cheshire, SK11 9FT, UK}
\affiliation[5]{Department of Chemistry, Sapienza University of Rome, P.le Aldo Moro 5, 00185 Rome, Italy}
\affiliation[6]{Departamento de Astronomía, Facultad Ciencias Físicas y Matemáticas, Universidad de Concepción \\Av. Esteban Iturra s/n Barrio
Universitario, Casilla 160, Concepción, Chile}
\affiliation[7]{Institute of Structure of Matter (ISM-CNR), Consiglio Nazionale delle Ricerche, 34149 Basovizza, \\Italy}
\affiliation[8]{Jodrell Bank Centre for Astrophysics, Department of Physics \& Astronomy, The University of \\Manchester, M13 9PL, U.K.}
\affiliation[9]{Niels Bohr Institute, University of Copenhagen, Jagtvej 155a, 2200 Copenhagen, Denmark}
\affiliation[10]{INAF - Istituto di Astrofisica e Planetologia Spaziali, via del Fosso del Cavaliere 100, 00133 \\Roma, Italy}
\affiliation[11]{Haystack Observatory, Massachusetts Institute of Technology, 99 Millstone Rd., Westford, MA \\01886, USA}

\emailAdd{andrea.bracco@obspm.fr}

\abstract{Magnetic fields play a central role in the star-formation process, from diffuse gas to 
the dense, starless,
molecular cloud cores that represent the first gravitationally bound structures on the path to star formation. Yet, the evolution of magnetic fields during this critical phase remains poorly understood. Recent studies suggest that cosmic-ray electrons interacting with magnetic fields in prestellar cores can produce detectable synchrotron emission at low radio frequencies, offering a novel probe of their magnetization in tandem with existing observational techniques. However, current instruments lack the angular resolution and sensitivity to exploit this signature. 
 
The Square Kilometre Array Observatory (SKAO) will provide the required capabilities enabling detections in nearby star-forming regions within a reasonable number of observation hours in AA* and AA4. Thanks to its large field of view, observations of low- to high-mass star-forming regions within the first kiloparsec from the Sun will enable both targeted studies of individual objects and statistical analyses over several hundreds of prestellar cores per pointing, marking a breakthrough in our understanding of their magnetic field properties.

This chapter outlines the scientific context, observational challenges, and prospects for probing magnetic fields in prestellar cores with SKAO, and highlights synergies with complementary facilities such as ALMA, as well as cross-disciplinary collaborations within the SKAO community.}

\begin{document}
\newcommand{\actaa}{Acta Astron.} 
\newcommand{\araa}{ARA\&A} 
\newcommand{\aar}{A\&ARv} 
\newcommand{\aapr}{A\&ARv} 
\newcommand{\ab}{Astrobiol.} 
\newcommand{\aj}{AJ} 
\newcommand{\apj}{ApJ} 
\newcommand{\apjl}{ApJL} 
\newcommand{\apjs}{ApJSS} 
\newcommand{\ao}{Appl. Opt.} 
\newcommand{\apss}{Astro. \& Space Sci.} 
\newcommand{\aap}{A\&A} 
\newcommand{\aaps}{A\&AS.} 
\newcommand{\baas}{Bull. Am. Astron. Soc.} 
\newcommand{\caa}{Chinese A\&A} 
\newcommand{\cjaa}{Chinese J. A\&A} 
\newcommand{\cqg}{Class. Quantum Gravity} 
\newcommand{\gal}{Galaxies} 
\newcommand{\gca}{Geo. Cosmo. Acta} 
\newcommand{\icarus}{Icarus} 
\newcommand{\jcap}{JCAP} 
\newcommand{\jgr}{J. Geophys. Res.} 
\newcommand{\jgrp}{J. Geophys. Res. Planets} 
\newcommand{\jqsrt}{J. Quant. Spectrosc. Radiat. Transf.} 
\newcommand{\memsai}{Mem. SAIt} 
\newcommand{\mnras}{MNRAS} 
\newcommand{\nat}{Nature} 
\newcommand{\nastro}{Nat. Astron.} 
\newcommand{\ncomms}{Nat. Commun.} 
\newcommand{\nphys}{Nat. Phys.} 
\newcommand{\na}{New Astron.} 
\newcommand{\nar}{New Astron. Rev.} 
\newcommand{\physrep}{Phys. Rep.} 
\newcommand{\pra}{Phys. Rev. A} 
\newcommand{\prb}{Phys. Rev. B} 
\newcommand{\prc}{Phys. Rev. C} 
\newcommand{\prd}{Phys. Rev. D} 
\newcommand{\pre}{Phys. Rev. E} 
\newcommand{\prl}{Phys. Rev. L.} 
\newcommand{\psj}{Planet. Sci. J.} 
\newcommand{\planss}{Planet. Space Sci.} 
\newcommand{\pnas}{Proc. Natl Acad. Sci. USA} 
\newcommand{\procspie}{Proc. SPIE} 
\newcommand{\pasa}{PASA} 
\newcommand{\pasj}{PASJ} 
\newcommand{\pasp}{PASP} 
\newcommand{\rmxaa}{RMXAA} 
\newcommand{\sci}{Science} 
\newcommand{\sciadv}{Sci. Adv.} 
\newcommand{\solphys}{Sol. Phys.} 
\newcommand{\sovast}{Soviet Ast.} 
\newcommand{\ssr}{Space Sci. Rev.} 
\newcommand{\uni}{Universe} 

\setlength{\bibsep}{0.0pt}  

\maketitle

\section{Introduction}\label{sec:intro}

The formation of solar-type stars results from a complex interplay of highly nonlinear physical mechanisms \citep[e.g.,][]{BallesterosParedes2020}. In the interstellar medium (ISM), turbulence and magnetic fields counteract the self-gravitational collapse of overdense regions, shaping the evolution of diffuse, multiphase gas into dense, filamentary molecular clouds. These environments provide the birthplace of stellar embryos—referred to as cores—where stars ultimately 
are born \citep{Hennebelle2012}. As these cores evolve, they transition from gravitationally unstable prestellar phases to main-sequence stars through successive protostellar stages, during which mass is steadily accreted from the surrounding gaseous envelope onto the central object, a protostar surrounded by an accretion disk \citep[e.g.,][]{Andre2000}.
\\
Magnetic fields play a key role in regulating gas flows across multiple spatial scales—from large molecular filaments \citep[e.g.,][]{PlanckXXXII,Soler2019b,Hacar2023} down to individual protostellar cores \citep[e.g.,][]{Maury2022}. The magnetic-field strength typically rises from a few $\mu$G in the diffuse ISM \citep[for column densities $N_{\rm H} < 5\times10^{21}$ cm$^{-2}$,][]{Heiles2003,Thompson2019} to mG in dense regions \citep[$N_{\rm H} \sim 10^{24}$ cm$^{-2}$,][]{Crutcher2010}. {In these environments, most cores appear to be only marginally supercritical—i.e., their masses exceed the magnetic support—making them prone to gravitational collapse \citep{Crutcher2012}. The relevant quantity is the mass-to-magnetic-flux ratio normalized to the critical value, $\mu = (M/\Phi_B)/(M/\Phi_B)_{\rm crit}$, where $(M/\Phi_B)_{\rm crit} \propto 1/\sqrt{G}$ and $G$ is the gravitational constant. The emergence of supercritical cores ($\mu > 1$) from subcritical ($\mu < 1$) diffuse molecular gas \citep{Hennebelle2019}} implies the operation of diffusion processes that partially decouple the magnetic field from gravity. Among these, ambipolar diffusion—arising from the relative motion between ions and neutrals in non-ideal magnetohydrodynamic (MHD) flows—has been proposed as a dominant mechanism \citep[e.g.,][]{Pinto2008,Momferratos2014}. Consequently, a key parameter governing star formation is the abundance of ionized particles that mediate the coupling between gas dynamics and magnetic fields in molecular clouds. While ultraviolet (UV) photons dominate ionization in the diffuse ISM, cosmic rays (CRs)—originating either from external or local sources—are expected to provide the main ionization channel in the UV-shielded interiors of molecular clouds \citep[e.g.,][]{Grenier2015,Padovani2020}. The overall CR ionization rate is primarily controlled by low-energy (MeV) protons \citep[e.g.,][]{Gabici2022}. Cosmic-ray electrons (CRe), though contributing to H$_2$ excitation  \citep{Ivlev2021,Padovani2022}, are predominantly expected to produce non-thermal synchrotron emission at GeV energies as they spiral around magnetic-field lines. 
\\
Observationally, this non-thermal radio emission was detected in extragalactic context \citep{Tabatabaei2025} but only in a few molecular clouds in our Galaxy at frequencies below 1 GHz, most notably in the diffuse Orion–Taurus ridge \citep{Bracco2023} and in compact regions surrounding the protostellar jet of DG Tau A \citep{FeeneyJohansson2019}. The absence of synchrotron emission from Galactic molecular clouds represents a long-standing problem that challenges our understanding of CRe and magnetic fields in star-forming regions. \\
In this chapter, we build upon the seminal works of \cite{Dickinson2015} and \cite{PadovaniGalli2018}, who proposed using the Square Kilometre Array Observatory (SKAO) to detect {\it in situ} synchrotron emission from dense molecular clouds. While other chapters in this volume discuss non-thermal processes in protostellar outflows and jets \citep[e.g.,][]{Sabatini2026} or at larger scales in molecular clouds \citep[e.g.,][]{Sun2026, Tahani2026}, here we focus on the potential of both the low- and high-frequency instruments—SKA-Low (50 - 350 MHz) and SKA-Mid (350 MHz - 15 GHz)—as unique probes of synchrotron emission in prestellar cores across low- to high-mass star-forming regions within the solar neighborhood ($\lesssim 1$~kpc). The SKAO will offer a transformative capability by combining exceptional sensitivity and angular resolution, enabling unprecedented statistical samples of dense starless cores and advancing our understanding of magnetic fields in star formation, particularly with the Array Assembly 4 (AA4) configuration \citep{AAsSKAO}.

This chapter is organized as follows: in Sect.~\ref{sec:state_art} we summarize the role of magnetic fields in dense starless cores. Section~\ref{sec:techniques} reviews current observational constraints on magnetic fields in dense cores using standard methods (dust polarimetry and Zeeman splitting). Section~\ref{sec:skao} introduces the low-frequency window and its relevance to {\it in situ} synchrotron emission, followed by the theoretical framework in Sect.~\ref{ssec:synch}. Observational forecasts with SKAO are presented in Sect.~\ref{ssec:skao}, including SKA-Low and SKA-Mid. Section~\ref{sec:synergy} outlines synergies—within SKAO and with ALMA—and Sect.~\ref{sec:end} provides the summary and conclusions.

\section{Magnetic fields in starless cores}\label{sec:state_art}

Establishing the dynamical evolution of prestellar cores is critical for understanding how mass is accreted during star formation. 
Observationally, in large surveys such as those conducted with the {\it Herschel} space observatory \citep[HGBS, HOBYS, Hi-GAL;][]{Andre2010, Russeil2019, Molinari2010}, the lack of direct measurements of the magnetic-field strength led to considering only the thermal pressure as the main agent counteracting the core's self-gravity.
In this framework, the dust-continuum-derived core mass, $m$, is compared to the Bonnor--Ebert mass, $m_\text{BE}$, which 
represents the minimum mass that a core must have for self-gravity to overcome internal pressure
\citep[see discussion in][]{Stutz2007}. 
\\
However, magnetic pressure can support a core against gravitational collapse even more effectively than thermal pressure \citep[see][]{Stutz2007}. It is therefore possible that many starless cores presently classified as unstable---that is, as collapsing---may in fact be magnetically supported and thus stable. This would have important implications, for example, for the derivation of the core mass function that sets the initial conditions of star formation \citep[e.g.,][]{Konyves2015}.
\\
The importance of the magnetic field in determining the stability of prestellar cores is characterized by the mass-to-(magnetic) flux ratio: if this ratio is below a certain critical value, that depends only on the gravitational constant, the magnetic field can prevent gravitational collapse. Observations of the Zeeman effect and polarized thermal dust emission (see Sect.~\ref{sec:techniques} for more details) suggest that the magnetic field is often strong enough to be dynamically important in cores, contributing to their support against collapse and influencing the star-formation process through magnetic braking \citep[e.g.,][]{Machida2020}. In particular, molecular cloud cores appear to be close to a magnetically critical state over a wide range of masses, with an important contribution from turbulent support, especially in the most massive cores \citep[e.g.,][and reference therein]{VazquezSemadeni2024}. 
However, it has been also shown that in some very massive regions the magnetic field strength required to slow down the global, parsec-scale collapse is higher than the maximum value expected in such regions \citep{Traficante18}. 
\\
Many uncertainties remain in this context due to the paucity of accurate magnetic-field strength measurements and the difficulty of reconstructing the three-dimensional field geometry. To date, it has not been possible to reconstruct the radial profile of the magnetic field in cloud cores, which would help constrain theoretical models. For a contracting cloud with a frozen-in magnetic field, its strength, $B$, is expected to scale roughly as a power of the density, $B \propto \rho^{\kappa}$, with $\kappa \approx 2/3$ ($\sim$0.67) if the contraction is isotropic \citep{Mestel66, Mestel77}. On the other hand, the ambipolar-diffusion-driven evolution of a magnetized, isothermal, self-gravitating cloud tends to produce a spatially uniform mass-to-flux ratio, resulting in a field-strength profile $B \propto r^{-1}$ \citep{LizanoShu89, LiShu96, CiolekBasu2000}. The extensive compilation of H\,\textsc{i}, OH, and CN Zeeman observations collected by \citet{Crutcher2012} suggests that the magnetic-field strength increases with density roughly as $B \propto n_{\rm H}^{\kappa}$ for $n_{\rm H} \gtrsim 300~\mathrm{cm}^{-3}$, with $\kappa = 0.65 \pm 0.05$ based on a Bayesian statistical analysis of the data. However, the interpretation of this relation and its statistical significance remain far from straightforward \citep[e.g.,][]{Whitworth25}. {Notably,  strong-field models predict a different behavior, with $\kappa \approx 0$ in the subcritical ISM and a transition to $\kappa \approx 0.5$ during the collapse of a spheroidal core \citep[e.g.,][and references therein]{Tritsis2015}.}
Additional observational constraints on the strength and structure of the magnetic field in starless cores are more needed than ever to advance our understanding of their physical properties. 

\section{Standard observational techniques}\label{sec:techniques}

Observational studies of magnetic fields in molecular clouds have so far relied on complementary tracers that probe different gas components. Dust polarization provides a measure of the orientation of the plane-of-sky magnetic field, $\vec{B}_{\perp}$ {(i.e., perpendicular to the line of sight)}, via the radiative-torque alignment of non-spherical dust grains with the ambient magnetic field \citep{HoangLazarian2016}, while Zeeman splitting yields the measure of the line-of-sight (LOS) component of the magnetic field, $B_{\rm LOS}$,
from the frequency separation of circularly polarized Zeeman-sensitive spectral components \citep{Crutcher2012}. 
\\
Dust polarization observations have revealed that magnetic fields are intricately linked to the filamentary structure of molecular clouds on all observable scales \citep{PlanckXXXV}. The large-scale orientation of the magnetic field transitions from being parallel to orthogonal to the elongated gas structures in diffuse and dense media, respectively, suggesting that magnetic tension plays a role in channeling material along field lines and assembling filaments that subsequently fragment and collapse \citep{Soler2017}. The column density at which this parallel-to-perpendicular transition occurs varies significantly from cloud to cloud, with a median value of $N_{\rm H} \approx 5\times10^{21}$~cm$^{-2}$ (or a visual extinction of $A_{\rm V} \approx 2.7$). This transition might correspond to the column density at which the magnetic field strength begins to increase with density in the Zeeman measurements compiled by \cite{Crutcher2012} (see Sect.~\ref{sec:state_art}), although the relation remains unclear. Moreover, on scales of $\lesssim 0.1$~pc, there is evidence that the orientation of the field changes again above $N_{\rm H} \approx 4 \times 10^{22}$~cm$^{-2}$ (or $A_{\rm V} \approx 21$), possibly due to gravity-induced flows \citep{Pillai20, Tapinassi24}, thereby calling into question the mechanisms of gas accretion toward dense cores.
For starless cores within filamentary structures in molecular clouds, resolving the internal magnetic morphology requires sub-parsec resolution, achievable so far only with stratospheric or ground-based millimetre and submillimetre telescopes. Ground-based polarimetric observations,  however, encounter significant challenges when probing both extremes of starless core column density. At low densities, the intrinsic emission is faint and is easily obscured by atmospheric noise, and at high densities, there is a considerable drop in grain-alignment efficiency due to the attenuation of the interstellar radiation field and lack of an internal source of radiation, confining the method’s sensitivity largely to an intermediate regime of column density \citep{Lin2024}. As a consequence, crucial information on how magnetic fields thread the transition from low-density to collapsing gas remains poorly constrained. Furthermore, the magnetic-field strengths inferred from dust polarization via the Davis–Chandrasekhar–Fermi (DCF) method are indirect and exhibit a wide dispersion for regions of comparable volume density, ranging from a few hundred~$\mu$G to several~mG \citep{Pattle2023}. These uncertainties are corroborated by synthetic polarization observations that test the efficacy of the DCF method in recovering true field strengths. Such studies reveal substantial issues, with the inferred values strongly dependent on the local Alfvénic conditions \citep[e.g.,][]{LiuDCF2021, Skalidis2021}.
\\
Zeeman splitting remains the only direct means of measuring interstellar magnetic-field strengths, {specifically the line-of-sight component}. In the cold, diffuse ISM, H \textsc{i} Zeeman detections reveal fields of a few~$\mu$G, which, in terms of magnetic pressure, are comparable to both the turbulent and thermal components along selected sight lines \citep{Heiles2005}. Zeeman observations of OH absorption and emission in molecular envelopes indicate fields of tens to roughly 100~$\mu$G, while CN emission in dense, active star-forming regions shows fields of several hundred~$\mu$G \citep{Falgarone08}. Zeeman measurements in starless cores remain exceedingly rare. In the massive cold clump G35.20w, a deep CN Zeeman experiment yielded an upper limit of $\sim$680~$\mu$G \citep{Pillai2016}. A recent FAST H \textsc{i} Zeeman detection in L1544 suggests $\lesssim$10~$\mu$G fields in the envelope of a low-mass starless core \citep{2022Natur.601...49C}, while CCS-based field-strength measurements toward another low-mass starless core, TMC-1C, indicate mG-level fields in its denser regions \citep{Koley2022}. Some of these observations have been interpreted to suggest that magnetic supercriticality may already arise within the envelopes of starless cores, with profound implications for star-formation timescales. However, the scarcity of such detections precludes the establishment of any meaningful—let alone statistically significant—trends. 

\section{The Low-frequency window}\label{sec:skao}

A promising and largely unexplored avenue to constrain the magnetization of dense starless cores arises from the low-frequency radio window. Complementing the traditional tracers described above (see Sect.~\ref{sec:techniques}), synchrotron emission provides an indirect probe of magnetic-field strengths through its dependence on the interactions between relativistic electrons and magnetic fields. In particular, the possibility of detecting non-thermal synchrotron 
emission of starless cores depends on the ability of Galactic CRe to penetrate the densest regions of molecular clouds, which in turn depends on the behavior of the magnetic field, since CR propagation can be influenced by magnetic focusing, mirroring, and scattering on Larmor-scale magnetic fluctuations
\citep{CesarskyVolk78,Chandran00,Padovani2011,Silsbee18,FitzAxen24}. 
\\
In the optically thin regime, synchrotron intensity decreases with frequency  as $I(\nu) \propto \nu^{\alpha}$, with $\alpha \sim -0.7$, making frequencies below 1 GHz particularly well-suited for detecting faint, diffuse emission from within molecular clouds. 
Despite this potential, remarkably few studies have attempted to detect {\it in situ} synchrotron emission directly arising from dense molecular gas \citep{Dickinson2015}. One of the earliest efforts was carried out by \cite{Jones2008}, who searched for leptonic emission from two nearby (3–4 kpc) molecular clouds using the Australia Telescope Compact Array (ATCA). Although no detection was made, their upper limits of $\approx0.5$\,mJy beam$^{-1}$ were consistent with an upper limit to the magnetic field strength of $500\,\mu$G. Follow-up observations toward Sgr B2 with similar techniques also yielded non-detections \citep{Jones2011}, which were interpreted as evidence for a suppression of $\sim$GeV CR diffusion into the dense cores of this cloud complex in the Galactic Centre \citep{Protheroe2008}.
\\
Perhaps the most compelling case of synchrotron emission emerging directly from within dense molecular material has been reported by \cite{Yusef-Zadeh2024}, who analyzed the Sgr B complex using multifrequency data at 1280, 333, 88, and 74\,MHz. They found a strong spatial correlation between steep-spectrum ($\alpha\approx-1$) radio continuum emission and dense molecular gas, inferring magnetic-field strengths of order 100 $\mu$G for gas densities $n_{\rm H}=10^4$–$10^5$ cm$^{-3}$.
A tentative detection of low-frequency synchrotron emission possibly associated with molecular material was also reported by \cite{Yusef-Zadeh2013} toward G0.13–0.13 at 74 MHz. However, the offset between the radio and CO emission peaks suggested that the observed emission did not originate directly from the dense gas itself. 
\\
In less extreme environments than the Galactic Centre, \cite{Bracco2023} detected synchrotron emission from the Orion–Taurus ridge, located at a distance of about 400 pc, on spatial scales of hundreds of parsecs, with magnetic-field strengths of a few tens of $\mu$G for gas column densities of $N_{\rm H_2} = 1.4\times10^{21}$ cm$^{-2}$.
On the core scale, \cite{Bracco2025} performed a statistical analysis using LOFAR 144 MHz data from the nearby Perseus molecular cloud \citep{Pezzuto2021}, analyzing more than 300 prestellar, dense cores. Using median-stacking techniques, they placed stringent statistical upper limits of 5 $\mu$Jy beam$^{-1}$, corresponding to magnetic-field strengths $\lesssim100\,\mu$G, among the tightest constraints to date for nearby star-forming regions.
Although detections remain scarce, these studies collectively highlight the emerging diagnostic power of low-frequency synchrotron observations for probing the magnetic environment of molecular clouds. With its unprecedented combination of sensitivity, angular resolution, and frequency coverage below 1 GHz, the SKAO will open a transformative window to study non-thermal emission from dense starless cores, the focus of this chapter. Such observations will provide critical, independent constraints on magnetic-field strengths and CR populations in the earliest phases of star formation, complementing and extending what is accessible through dust polarization and Zeeman measurements.

\subsection{Basics and assumptions for synchrotron emission}\label{ssec:synch}

Galactic synchrotron emission arises from the interaction between interstellar magnetic fields and CRe \citep[e.g.,][]{GinzburgSyrovatskii1964}. These relativistic particles gyrate around magnetic field lines, emitting non-thermal radiation that depends on their energy, $E$, and on the strength of the perpendicular magnetic-field component, $B_{\perp} = |\vec{B}_{\perp}|$. The total synchrotron emissivity, expressed in units of power per unit volume, frequency ($\nu$), and solid angle, is the sum of the emissivities linearly polarized parallel ($\varepsilon_{\nu,\parallel}$) and perpendicular ($\varepsilon_{\nu,\perp}$) to $\vec{B}_{\perp}$:

\begin{eqnarray}\label{eq:epsnu}
\varepsilon_{\nu,\|}(\Vr) &=& \int_{m_{e}c^{2}}^{\infty}\frac{j_{e}(E)}{\varv_{e}}P_{\nu,\|}^{\rm em}(E,{\bperp}(\Vr))\,\ud E,\\\nonumber
\varepsilon_{\nu,\perp}(\Vr) &=& \int_{m_{e}c^{2}}^{\infty}\frac{j_{e}(E)}{\varv_{e}}P_{\nu,\perp}^{\rm em}(E,{\bperp}(\Vr))\,\ud E,
\end{eqnarray}

where $j_e(E)$ is the CRe energy spectrum, $\varv_e$ is the electron velocity, $m_e$ is the electron mass, $c$ is the speed of light, and $\Vr$ denotes position.
\\
The quantities
\begin{eqnarray}
\label{power}
P_{\nu,\|}^{\rm em}(E,\Vr)&=&\frac{\sqrt{3}e^{3}}{2m_{e}c^{2}} B_{\perp}(\Vr) [F(x)-G(x)],\\\nonumber
P_{\nu,\perp}^{\rm em}(E,\Vr)&=&\frac{\sqrt{3}e^{3}}{2m_{e}c^{2}} B_{\perp}(\Vr) [F(x)+G(x)],
\end{eqnarray}
represent the power per unit frequency emitted at frequency $\nu$ in the two polarization directions \citep[see also][]{Longair2011,Padovani2021}. Here, $e$ is the elementary charge, $x=\nu/\nu_c$, and $\nu_c$ is the critical frequency defined as

\begin{equation}\label{nuc}
\nu_c[B_{\perp}(\Vr),E]=\frac{3eB_{\perp}(\Vr)}{4\pi m_{e}c}\left(\frac{E}{m_e c^2}\right)^{2}.
\end{equation}
\\
The functions $F(x)$ and $G(x)$ are given by
\begin{equation}
F(x)=x\int_{x}^{\infty}K_{5/3}(\xi)\,\ud\xi
\qquad
\mbox{and}
\qquad
G(x)=x\, K_{2/3}(x),
\end{equation}
where $K_{5/3}$ and $K_{2/3}$ are the modified Bessel functions of the second kind of orders 5/3 and 2/3, respectively.
By integrating Eqs.~\ref{eq:epsnu} along the LOS and accounting for the telescope beam ($\Theta_{\rm T}$), one obtains the synchrotron-specific Stokes $I$ as:
\begin{equation}\label{eq:tb}
    I(\nu) = \Theta_{\rm T} \circledast \int_{\rm LOS} [\varepsilon_{\nu,\|} + \varepsilon_{\nu,\perp}]\,\ud r,
\end{equation}
where the symbol $\circledast$ denotes convolution with a Gaussian-approximated beam $\Theta_{\rm T}$ characterized by its full width at half maximum (FWHM), $\theta_{\rm T}$. In the following, to make predictions of the Stokes $I$ brightness for the SKAO, we adopt both $\theta_{\rm T} = 10''$ and $\theta_{\rm T} = 5''$, for SKA-Low and -Mid, respectively. The methodology described above is analogous to that applied in \cite{Padovani2018, Padovani2021} and \cite{Bracco2022,Bracco2023, Bracco2024, Bracco2025}.
\\
Models and simulations developed to interpret Galactic synchrotron emission generally assume that the CRe populations contributing to the emission above and below 408~MHz follow power-law energy spectra $j_e(E)\propto E^s$ with exponents $s = -3$ and $s = -2$, respectively \citep{Sun2008,Waelkens2009,Reissl2019,Wang2020}. However, as CRe propagate through the ISM, they lose energy via interactions with matter, magnetic fields, and radiation. These processes deplete the population of relativistic electrons, thereby modifying the initial energy spectrum.
\\
Significant progress in determining the high-energy spectrum of Galactic CRs has been achieved through observations by the \emph{Fermi} Large Area Telescope \citep[\emph{Fermi} LAT;][]{Ackermann2010}, the balloon-borne PAMELA experiment \citep{Adriani2011}, and the Alpha Magnetic Spectrometer (AMS-02) onboard the International Space Station \citep{Aguilar2014}. More recently, the two \emph{Voyager} probes have provided crucial data on the low-energy spectrum, down to $\simeq 3~\mathrm{MeV}$, largely unaffected by solar modulation after crossing the heliopause \citep{Cummings2016,Stone2019}. 
\\
In this chapter, rather than assuming power-law energy spectra, we adopt $j_e(E)$ derived from data-driven approaches that incorporate recent direct measurements. Nevertheless, a comprehensive understanding of the origin and propagation of Galactic CRs remains elusive, due to degeneracies among key parameters and uncertainties related to re-acceleration, convection, and the diffusion coefficient \citep[e.g.,][]{Grenier2015}. Finally, an important caveat of our method is that the synthetic models include only an external population of CRe incident on the cloud, with no internal source of acceleration {that can be found in active star forming regions \citep{Pineda2024}}.
 
\subsection{Forecasts with SKAO}\label{ssec:skao}
In this section, we use numerical models of prestellar cores from ideal MHD simulations of molecular clouds to present observational forecasts with SKA-Low and SKA-Mid. The synthetic models are post-processed following the methodology described in Sect.~\ref{ssec:synch}, and the time estimates for both AA* and AA4 are derived using the SKAO sensitivity calculator\footnote{\url{http://sensitivity-calculator.skao.int/}} \citep[e.g.,][for the low frequency part]{Sokolowski2022}. Compared to previous studies limited to analytical magnetostatic models of magnetized cores \citep[e.g.,][]{Bracco2025}, this work provides, for the first time, estimates of the synchrotron intensity of simulated cores embedded within their parent molecular cloud. 
\begin{wrapfigure}[19]{r}[0pt]{0.5\textwidth}
  \vspace{0pt}
  \centering
  \includegraphics[width=0.5\textwidth]{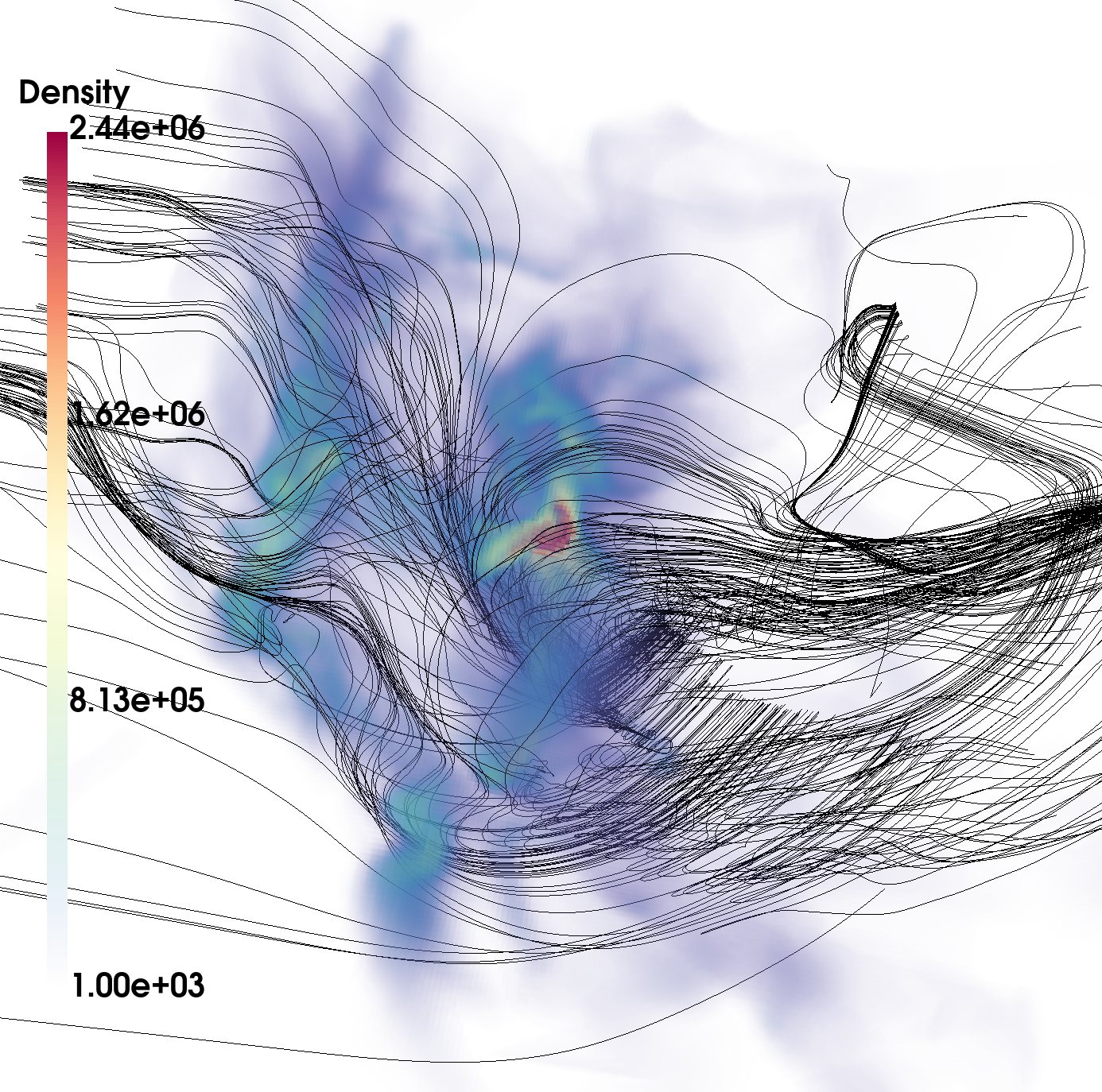} 
  \caption{\it 3D rendering of the {subcritical} MHD case, where the density field in units of cm$^{-3}$ is shown in colors with black magnetic-field lines including the host molecular cloud and the embedded core.}
  \label{fig:3D_render}
\end{wrapfigure}
The MHD simulations are isothermal and consist of 3D cubes made with adaptive mesh refinement using {\tt RAMSES} \citep{Teyssier2002,Fromang2006} to model star formation in molecular clouds \citep[see][]{Haugbolle2018,Kuffmeier2023}. 
The simulated cubes are periodic boxes with length of 4 pc.
The total mass of the molecular cloud is 3000 M$_\odot$ and the mean magnetic field strength is 7.2\,$\mu$G. Turbulence is sinusoidally driven for 10 crossing times of the gas with root-mean-square velocity of $v_\mathrm{RMS}\sim1.8$ km s$^{-1}$ before self-gravity is turned on and stars, modeled as sink particles, are allowed to form.
To estimate the observational tracers of magnetic fields in a prestellar core, we excise a cubical region of $(0.5~ \mathrm{pc})^3$ in volume, a few thousand years before the formation of a star that has accreted about $1.7~ M_{\odot}$ by the end of the simulation. 
The densest parts of the prestellar core are resolved with 25 au in the original simulations and we convert the excised region of $(0.5~ \mathrm{pc})^3$ to a cubical grid consisting of $64^3$ cells with uniform resolution. We use two reference cases depending on the magnetic-field strength averaged in $10^4$ au from the center of mass of each core. {The first case corresponds to a magnetic-field strength of $90\,\mu$G and a mass-to-flux ratio $\mu \approx 0.8$ (subcritical case, Sect.~\ref{sec:intro}), while the second case corresponds to $30\,\mu$G and $\mu \approx 1.5$ (supercritical case).} In Fig.~\ref{fig:3D_render}, we show the 3D rendering\footnote{The rendering was produced with {\tt PyVista}, \url{http://docs.pyvista.org/index.html}.} of the density (in colors) and magnetic fields (black lines) of the {subcritical} case. The core can be seen at the center of the plot surrounded by the turbulent magnetic field of the host molecular cloud.

\subsubsection{The case of SKA-Low}\label{sssec:skaLow}
\begin{figure}[t]
    \centering
	\includegraphics[width=1\columnwidth]{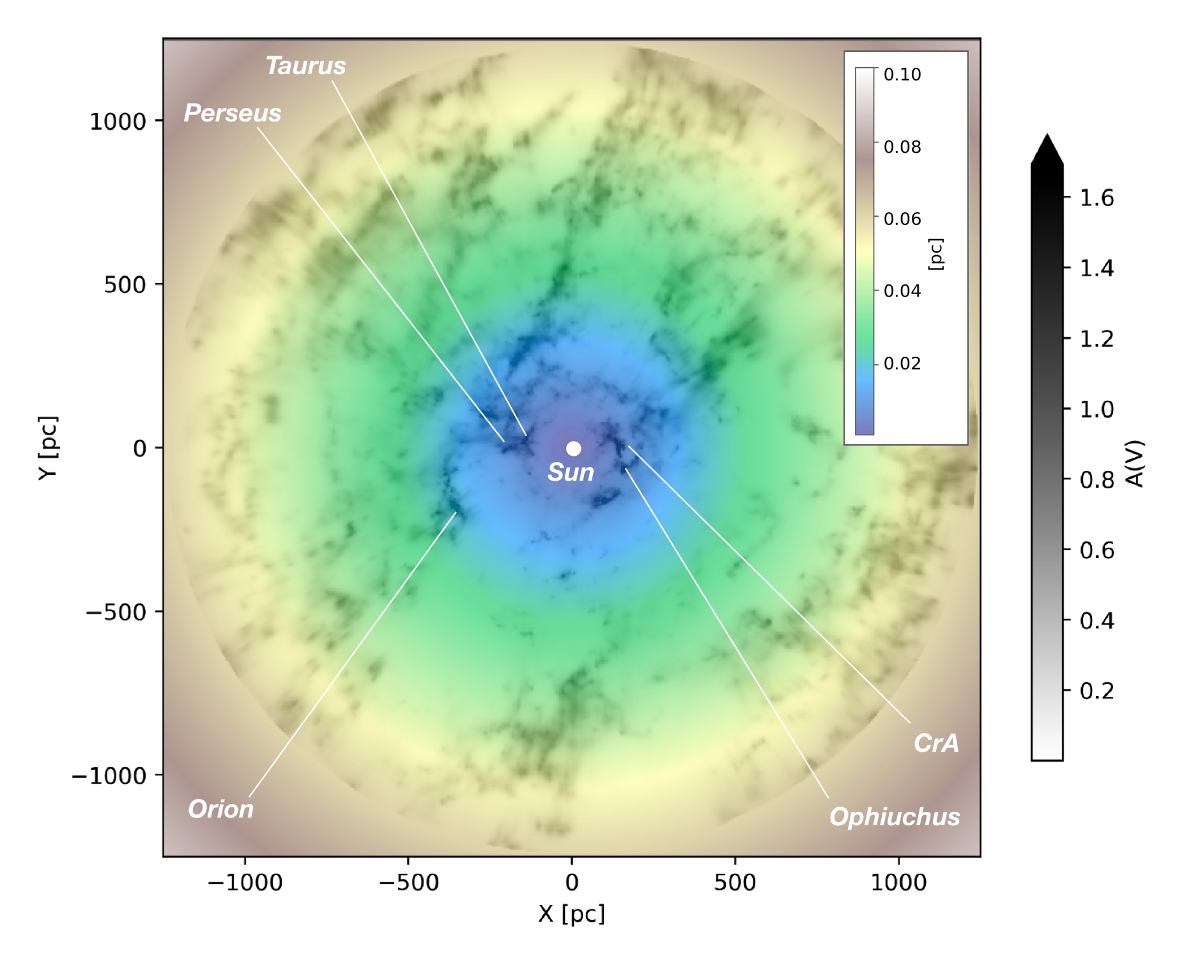}
    \caption{\it SKA-Low physical resolution in pc (a Gaussian beam of 10" FWHM is considered) overlaid on a bird's-eye view of the 3D dust extinction map around the Sun adapted from \cite{Edenhofer2024}. The Galactic center is toward the right side of the plot. A few molecular clouds listed in Table~\ref{tab:MHD_emb_models} are labeled in white: Orion, Perseus, Taurus, Corona Australis (CrA), and Ophiuchus.}
    \label{fig:resol}
\end{figure}
Thanks to the wide range of angular-scale sensitivity of SKA-Low--from several degrees down to 10"--the instrument will offer a breakthrough opportunity to probe the multi-scale magnetic-field structure of molecular clouds, reaching the physical scales of individual prestellar cores ($\sim 0.1$~pc). This capability will be transformative for conducting statistical studies of core properties across most star-forming regions in the solar neighborhood and beyond. As shown in Fig.~\ref{fig:resol}, which provides a bird’s-eye view of $A_{\rm V}$ around the Sun from \cite{Edenhofer2024}, the highest angular resolution of SKA-Low will allow us to spatially resolve cores in dense clouds ($A_{\rm V} > 1$) within 500 pc of the Sun—reaching a physical resolution of about 0.01 pc—and to marginally resolve them out to 1 kpc. Although the present forecasts focus on the best-known nearby star-forming molecular clouds listed in Table~\ref{tab:MHD_emb_models} and labeled in Fig.~\ref{fig:resol}, we note that SKA-Low may also serve as a powerful probe of prestellar-core properties in more distant, high-mass star-forming regions.
\begin{figure}[t]
    \centering
\includegraphics[width=1\columnwidth]{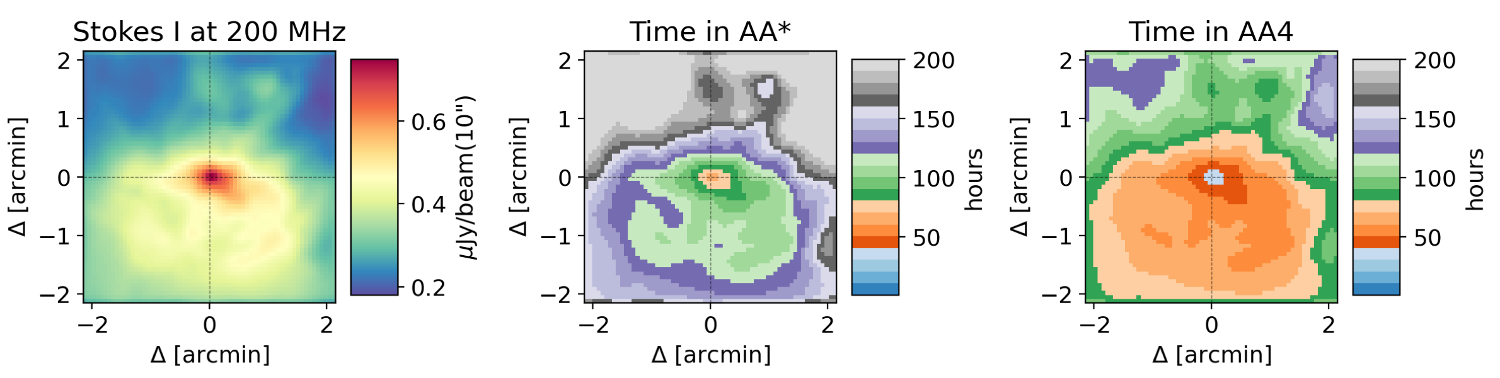}
    \caption{\it Time estimates to reach a signal-to-noise ratio of five at 200 MHz within a 300 MHz bandwidth in both AA* and AA4 configurations using synthetic data from MHD simulations. As an example, we show a simulated core embedded in one Orion-like cloud in the {subcritical} case (see Table~\ref{tab:MHD_emb_models}).}
\label{fig:MHD_mod_Emb_StrongB}
\end{figure}
\\
We produced synthetic models of synchrotron emission at 200 MHz using the {subcritical} and supercritical reference cases with various choices of $j_e$ in Eq.~\ref{eq:epsnu}. We considered both the tabulated $j_e$ spectrum from \citet{Orlando18}, which represents the interstellar CRe energy distribution at the Sun’s location after propagation through a modeled Galactic-scale magnetic-field structure, and the analytical Model A of \citet{Bracco2024}, which was used to empirically reproduce the observed radio spectral index of \citet{Guzman2011} at intermediate Galactic latitudes. We notice that in our approach we have neglected the possible extinction effects of radio photons, such as synchrotron self-absorption and free-free absorption \citep[e.g.,][]{Rybicki1979}. Both effects are not relevant for the kind of interstellar environment we focus on, as prestellar cores are sources with low ionization fraction and rather low magnetic-field strengths.\\
As an example, the left panel of Fig.~\ref{fig:MHD_mod_Emb_StrongB} shows the synthetic Stokes $I$ map at 200 MHz, convolved with a 10" FWHM Gaussian beam, for the {subcritical} case using the $j_e$ spectrum from \citet{Orlando18} at a distance of 400 pc—corresponding to an Orion-like cloud (see Fig.~\ref{fig:resol}). To mitigate the influence of specific magnetic-field configurations within the cores, Stokes $I$ estimates are calculated statistically by averaging the synthetic brightness over all principal axes of the simulated cubes.\\
In our synthetic maps, no instrumental effects are included apart from Gaussian smoothing. However, SKA-Low should be sufficiently complete to recover diffuse emission on the few-arcminute scales shown in the figure.\\
The central and right panels of Fig.~\ref{fig:MHD_mod_Emb_StrongB} display the estimated observing times for the AA* and AA4 configurations, respectively. These integration times correspond to a signal-to-noise ratio of 5 at 200 MHz within the full 300 MHz bandwidth of SKA-Low, assuming a minimum elevation of 30$^{\circ}$ toward the Orion position listed in Table~\ref{tab:MHD_emb_models}. Briggs weighting was considered in the sensitivity calculator (robust value =
0). The models predict that a single {subcritical} dense starless core in Orion could be detected at its center in fewer than 100 hours with AA*, and in less than 50 hours with AA4. Thanks to the wide field of view of SKA-Low (on the order of 10$^{\circ}$), these integration times could be significantly reduced if, instead of targeting a single object, one considers a statistical detection through median stacking, as recently demonstrated by \citet{Bracco2025} using LOFAR. In this case, the observation time required would decrease by a factor of $\sqrt{N_{\rm c}}$, where $N_{\rm c}$ denotes the number of known prestellar cores in a given cloud. In nearby molecular clouds, such as those listed in Table~\ref{tab:MHD_emb_models}, the number of identified prestellar cores ranges from a few hundred to several hundred, as revealed by infrared facilities such as the {\it Herschel} space observatory \citep[e.g.,][]{Kirk2013,Bresnahan2018,Konyves2020,Ladjelate2020,Pezzuto2021,Kirk2024}.\\
Table~\ref{tab:MHD_emb_models} summarizes the estimated integration times at the peak of the cores for the main star-forming regions in the solar neighborhood. Depending on sky location and the adopted model, the required observing times range from a few tens to about a thousand hours (in bold text in the table) for the detection of a single prestellar core. We emphasize that more massive star-forming clouds at greater distances in the inner Galaxy—despite their poorer angular resolution—may also be suitable targets, possibly exhibiting magnetic field strengths exceeding $\sim$100~$\mu$G in extended prestellar-core complexes \citep[e.g.,][]{Pillai2015,Redaelli2021,Pillai2023}.\\
If detection of the prestellar-core peak brightness is feasible in a few molecular clouds using the $\sim$300 SKA-Low stations in the AA* configuration within a reasonable observation time, then accessing the greater sensitivity of AA4—with more than 500 stations—will be essential. The AA4 configuration would increase both the statistical sample of detectable sources and improve efficiency by about 100\%, corresponding to a twofold reduction in observation time (e.g., from $\lesssim 100$ hours in AA* to $\lesssim 50$ hours in AA4 toward Orion, see Table~\ref{tab:MHD_emb_models}).  

\begin{sidewaystable}[ht]
	\centering
	\caption{\it
    SKA-Low time estimates in AA* and AA4 configurations to detect one simulated core in the {subcritical (sub) and supercritical (super)} cases \citep{Kuffmeier2023} with a signal-to-noise ratio of 5 at 200 MHz within a 300 MHz bandwidth. Briggs weighting was considered in the sensitivity calculator (robust value = 0). The tabulated sensitivities are estimated for 1h observation and minimum elevation of 20$^{\circ}$. The CRe energy spectrum from \citet{Orlando18} was used in the numerical models. In parenthesis, the time estimates for the CRe energy spectrum taken from \cite{Bracco2024} are also shown.}
	\label{tab:MHD_emb_models}
	\begin{tabular}{|l|cccc|cccc|} 
		\hline
		Progenitor cloud & {\small RA, Dec} & Distance & Sensitivity & Sensitivity & Time in AA* & Time in AA4 & Time in AA* & Time in AA4\\
          & [deg] & [pc] & AA* & AA4 & sub & sub & super & super\\
          &  & & [$\mu$Jy/beam(10")] & [$\mu$Jy/beam(10")]& [h] & [h] & [h] & [h]\\
		\hline
		Orion & 84, -7 & 400 & 9 & 5 & 59(175) & 32(95) &124(343) & 67(187)\\
		Ophiuchus & 248, -25 & 140 & 16 & 10 & 90(261) & 54(156) & 194(524) & 117(314)\\
        Lupus & 236, -34 & 190 & 17 & 10 & 99(291) & 57(166) & 212(580) & 121(331)\\
        Corona Australis & 286, -37 & 150 & 18 & 10 & 101(291) & 59(172) & 215(582) & 127(344)\\
        Taurus & 66, +25 & 140 & 23 & 14 & 132(379) & 77(222) & 282(760) & 165(446)\\
        Musca & 186, -71 & 150 & 26 & 15 & 151(436) & 87(251) & 323(873) & 186(502)\\
        Chamaeleon & 180, -78 & 185 & 29 & 17 & 170(497) & 99(290) & 362(989) & 211(577)\\
        Serpens & 279, 0 & 480 & 34 & 20 & 232(693) & 138(412) & 490({\bf 1349}) & 292(803)\\
        Pipe & 259, -27 & 130 & 36 & 21 & 204(584) & 118(337) & 436({\bf 1171}) & 252(677)\\
        Aquila & 277.5, -2 & 133 & 38 & 22 & 213(611) & 127(364) & 456({\bf 1226}) & 271(730)\\
        Perseus & 54, +31 & 300 & 55 & 33 & 351({\bf 1053}) & 209(626) & 744({\bf 2068}) & 442({\bf 1229})\\
		\hline
	\end{tabular}
\end{sidewaystable}

\begin{sidewaystable}[ht]
	\centering
	\caption{\it
    SKA-Mid Band-1 time estimates in AA* and AA4 configurations to detect one simulated core in the {subcritical (sub) and supercritical (super)} cases \citep{Kuffmeier2023} with a signal-to-noise ratio of 5 at 800 MHz within a 435 MHz bandwidth. Briggs weighting was considered in the sensitivity calculator (robust value = 0). The tabulated sensitivities are estimated for 1h observation and minimum elevation of $45^{\circ}$. The CRe energy spectrum from \citet{Orlando18} was used in the numerical models. In parenthesis, the time estimates for the CRe energy spectrum taken from \cite{Bracco2024} are also shown.}
	\label{tab:MHD_emb_models_Mid1}
	\begin{tabular}{|l|cc|cccc|} 
		\hline
		Progenitor cloud  & Sensitivity & Sensitivity & Time in AA* & Time in AA4 & Time in AA* & Time in AA4\\
          &  AA* & AA4 & sub & sub & super & super\\
          &  [nJy/beam(5")] & [nJy/beam(5")]& [h] & [h] & [h] & [h]\\
		\hline
		Orion &  2160 & 506 & 116(378) & 27(89) & 219(653) & 51(153)\\
		Ophiuchus & 2400 & 560 & 121(377) & 28(88) & 229(662) & 54(155)\\
        Lupus & 2400 & 560 & 121(378) & 28(88) & 229(662) & 54(155)\\
        Corona Australis & 2400 & 560 & 121(377) & 28(88) & 229(662) & 54(155)\\
        Taurus & 3700 & 780 & 186(582) & 39(123) & 354(1021) & 75(215)\\
        Musca & 3040 & 780 & 153(478) & 39(123) & 290(839) & 75(215)\\
        Chamaeleon & 3150 & 1000 & 159(496) & 50(157) & 301(869) & 96(276)\\
        Serpens & 4060 & 800 & 226(747) & 45(147) & 424(1278) & 83(252)\\
        Pipe & 3500 & 720 & 176(550) & 36(113) & 334(966) & 69(199)\\
        Aquila & 4200 & 840 & 212(661) & 42(132) & 401({\bf 1159}) & 80(232)\\
        Perseus & 3800 & 1350 & 195(618) & 69(220) & 370({\bf 1079}) & 131(383)\\
		\hline
	\end{tabular}
\end{sidewaystable}

\begin{sidewaystable}[ht]
	\centering
	\caption{\it
    SKA-Mid Band-2 time estimates in AA* and AA4 configurations to detect one simulated core in the {subcritical (sub) and supercritical (super)} cases \citep{Kuffmeier2023} with a signal-to-noise ratio of 5 at 1.31 GHz within a 720 MHz bandwidth. Briggs weighting was considered in the sensitivity calculator (robust value = 0). The tabulated sensitivities are estimated for 1h observation and minimum elevation of 45$^{\circ}$. The CRe energy spectrum from \citet{Orlando18} was used in the numerical models. In parenthesis, the time estimates for the CRe energy spectrum taken from \cite{Bracco2024} are also shown.}
	\label{tab:MHD_emb_models_Mid2}
	\begin{tabular}{|l|cc|cccc|} 
		\hline
		Progenitor cloud  & Sensitivity & Sensitivity & Time in AA* & Time in AA4 & Time in AA* & Time in AA4\\
          &  AA* & AA4 & sub & sub & super & super\\
          &  [nJy/beam(5")] & [nJy/beam(5")]& [h] & [h] & [h] & [h]\\
		\hline
		Orion &  680 & 187 & 51(174) & 14(48) & 91(286) & 25(79)\\
		Ophiuchus & 720 & 168 & 50(164) & 12(38) & 91(274) & 21(64)\\
        Lupus & 720 & 168 & 50(164) & 12(38) & 91(274) & 21(64)\\
        Corona Australis & 720 & 168 & 50(164) & 12(38) & 91(274) & 21(64)\\
        Taurus & 1080 & 288 & 76(246) & 20(65) & 136(411) & 36(110)\\
        Musca & 828 & 216 & 58(188) & 15(49) & 104(315) & 27(82)\\
        Chamaeleon & 884 & 240 & 62(201) & 17(55) & 112(337) & 30(91)\\
        Serpens & 840 & 240 & 65(226) & 19(65) & 116(369) & 33(106)\\
        Pipe & 800 & 196 & 56(182) & 14(45) & 101(305) & 25(75)\\
        Aquila & 880 & 240 & 62(200) & 17(55) & 111(335) & 30(91)\\
        Perseus & 1400 & 364 & 100(330) & 26(86) & 180(550) & 47(143)\\
		\hline
	\end{tabular}
\end{sidewaystable}

\subsubsection{The case of SKA-Mid}\label{sssec:skaMid}

The same conclusions drawn for SKA-Low also apply to SKA-Mid in Bands 1 and 2, as shown in Tables~\ref{tab:MHD_emb_models_Mid1} and~\ref{tab:MHD_emb_models_Mid2}. The time estimates for SKA-Mid\footnote{We do not consider Band 5 because non-thermal synchrotron emission may be contaminated by thermal dust continuum above 10 GHz.} were calculated assuming the full available bandwidths centered at 800 MHz and 1.3 GHz, respectively. We adopted Briggs weighting (robust parameter = 0) and imposed a minimum elevation of 45$^{\circ}$.\\
At an angular resolution of 5", Band 2 offers the most promising scenario, with individual prestellar-core detections at the brightness peak requiring between a few tens and a few hundred hours of integration in both AA* and AA4 configurations, regardless of model variability. The adopted angular resolution was chosen to allow for a uniform comparison among the time estimates for different molecular clouds. We note, however, that for specific clouds, the angular resolution achievable with SKA-Mid may reach sub-arcsecond scales, providing unprecedented detail of the sources. Owing to this capability, SKA-Mid may also enable the study of starless cores located beyond 1 kpc from the Sun, improving the angular resolution by at least a factor of two compared with SKA-Low (see Fig.~\ref{fig:resol}), and allowing prestellar-core investigations in more massive star-forming regions than those found in the solar neighborhood.\\
However, the most valuable advancement provided by the SKAO will be the potential combination of its two instruments, enabling unmatched constraints on the spectral dependence of synchrotron emission across nearly two orders of magnitude in frequency. This will represent a major breakthrough in the study of CRe properties and magnetic fields during the earliest stages of the star-formation process.

\section{Synergies}\label{sec:synergy}

In this section, we highlight how the study presented in this chapter connects with other SKAO science cases in this volume and with complementary facilities, especially the Atacama Large Millimeter Array (ALMA).

\subsection{Synergies within SKAO}\label{ssec:skaomid}
Observations targeting prestellar cores in nearby molecular clouds will also provide essential data for several complementary scientific cases, such as {Zeeman measurements of magnetic field strengths (see the dedicated chapter by \citet{Bourke2026} in this volume) and
} the study of radio emission from Young Stellar Objects (YSOs) and outflows in regions of active star formation like Orion (see by \citet{Sabatini2026} in this volume).
Using the Orion case reported in Tables~\ref{tab:MHD_emb_models_Mid1} and~\ref{tab:MHD_emb_models_Mid2} as a reference, the native angular resolution of SKA-Mid---i.e. $\sim 1.3'' \times 1.2''$ in Band~1 and $\sim 0.8'' \times 0.7''$ in Band~2---will provide, in one hour of integration time, a sensitivity of $\sim 2.5~\mu$Jy~beam$^{-1}$ in Band~1 and $\sim1.2~\mu$Jy~beam$^{-1}$ in Band~2. In commensality with the prestellar core science cases, the sensitivity obtained at $\sim 1''$, combined with the remarkable SKA-Mid field of view (e.g. $\sim 100'$ in Band~1 and $\sim60'$ in Band~2; see \href{https://www.skao.int/sites/default/files/documents/Technical%20Information%20Summary%20Sheet%20SKA-AA4.pdf}{SKA-AA4 Technical Summary}), will enable detailed studies of YSO jet properties and kinematics within the same star-forming regions.
\\
Typical flux densities of radio jets and knots in HH objects at centimetre wavelengths ($\sim 10$~GHz) are in the range $\sim 0.1$ to a few mJy (e.g. \citealt{Anglada2018}). The emission is expected to be substantially higher for non-thermal sources at lower frequencies. Assuming a spectral index of $-0.6$ (e.g. \citealt{Osorio2017, Vig2018}), a source with a flux density of 0.2~mJy at 10~GHz is expected to reach $\sim 1$~mJy at the frequencies covered by SKA-Mid Bands~1 and~2. For a conservative polarisation fraction of 5--10\%, we estimate the expected polarised flux density of a non-thermal radio jet to be on the order of 50--100~$\mu$Jy.
\\
This synergistic approach is crucial for two main objectives: first, to extend continuum studies of protostellar jets beyond the single case studied so far (i.e., HH 80–81; \citealt{RodriguezK2025}); and second, to use high angular resolution for proper-motion studies of these jets, thereby enabling routine 3D velocity measurements of faint thermal and non-thermal knots, specifically targeting structures much fainter than the few mJy objects typically detected by current facilities.

\subsection{Synergies with ALMA}\label{ssec:skaoalma}

The low-frequency observations of dense prestellar cores proposed in this chapter will enable powerful multi-wavelength synergies with ALMA's chemical and thermal diagnostics of molecular clouds. The combination of SKAO and ALMA will provide complementary insights into the intertwined effects of magnetization and ionization on the physical and chemical properties of dense starless cores (see Sect.~\ref{sec:intro}).
\\
{Beyond the use of ALMA for high-resolution dust-polarization measurements (thanks to the future wideband sensitivity upgrade of ALMA), maps of \mbox{HCN} and its protonated form \mbox{HCNH$^+$}---together with isotopologues such as H$^{13}$CN and HC$^{15}$N---provide spatially resolved constraints on the ionisation fraction and the chemical clock of cold, dense gas \citep[e.g.,][]{Quenard2017,Fontani2021}.}
The abundance ratio [HCNH$^+$]/[HCN] is particularly sensitive to the local ionisation environment, since HCNH$^+$ forms via proton-transfer sequences initiated by H$_3^+$ and is removed through dissociative recombination, whose rates have now been more firmly constrained \citep{Ayouz2024}. Moreover, a new method based on ALMA observations of the ortho form of H$_2$D$^+$, CO, DCO$^+$, and HCO$^+$ (or their most likely optically thin isotopologues) have recently been proposed as an excellent proxy for the CR ionisation rate in cold, dense prestellar regions \citep{Bovino20, Sabatini20, Sabatini23, Redaelli24, Redaelli25}, from which the electron abundance can be derived (e.g. \citealt{Latrille25}). Similar methodologies can be applied to more active star forming regions like Orion \citep[e.g.,][]{Socci2024}.
\\
In parallel, SKA-Low and -Mid can constrain the magnetic and relativistic-electron context that mediates this chemistry. Low-frequency synchrotron maps (and their spectral slope) trace the product of CRe density and $B_\perp$, while SKA-Mid enables additional magnetic diagnostics (e.g., prospects for Zeeman-sensitive species from \citet{Bourke2026} in this volume) and robust separation of thermal versus non-thermal continua across hundreds of MHz to GHz. 
\\
Cross-comparison with ALMA ions such as HCO$^+$, N$_2$H$^+$, and DCO$^+$ can then quantify ion–neutral drift and the efficiency of magnetic-field diffusion processes such as ambipolar diffusion. These tracers link chemical stratification to magnetic support and to the neutral–ion coupling that regulates collapse on core scales. Recent theoretical work by \citet{Tritsis2023} further suggests that HCNH$^+$ is among the most promising molecular ions to detect such ion–neutral drift signatures, providing a strong motivation for future joint SKA–ALMA studies. 
\\
In practice, ALMA establishes the thermal and chemical baselines (density, temperature, depletion, deuteration), while SKAO constrains the non-thermal and magnetic boundary conditions. Furthermore, the extremely high angular resolution achievable with ALMA — reaching sub-arcsecond scales even for the lowest rotational transitions, such as HCO$^+$ (1–0) and N$_2$H$^+$ (1–0) in the 3 mm band — will enable us to accurately study the spatial extent of the ionized emission regions and their immediate surroundings. Together, SKAO and ALMA will close the loop between ionization chemistry and MHD in prestellar cores.

\section{Summary and conclusion}\label{sec:end}
In this chapter, we have shown that the SKAO will consolidate, in an unprecedented manner, the low-frequency radio window for probing the magnetization of dense starless cores through {\it in situ} low-frequency synchrotron emission. Using MHD-based synthetic observations and SKAO sensitivities, we find that individual cores in nearby clouds ($\lesssim 1$~kpc) are within reach—especially in the AA4 configuration—and that the wide fields of SKA-Low and SKA-Mid enable statistical detections through stacking techniques. Together, SKA-Low and SKA-Mid provide surface-brightness and spectral-index constraints that offer unique insights into the magnetization of dense starless cores, complementing dust polarimetry and Zeeman measurements.
\\
We emphasize that a non-detection of synchrotron emission from dense cores with the SKAO would truly challenge our understanding of their cosmic-ray and magnetic-field properties. A key synergy will also emerge with ALMA: chemical maps of HCN, HCNH$^+$, and their isotopologues, together with HCO$^+$, N$_2$H$^+$, and DCO$^+$, will quantify ionization and chemistry. Joint analyses will thus enable estimates of the cosmic-ray ionization rate and tests of ion–neutral coupling (e.g., ambipolar diffusion) on core scales. In short, the SKAO adds the missing non-thermal component to the multi-wavelength picture, enabling a coherent view of chemistry, ionization, and magnetic fields in the earliest stages of star formation.

\section*{Acknowledgments}
{\small AB and DG acknowledge financial support from the INAF initiative ``IAF Astronomy Fellowships in Italy'' (grant name MEGASKAT) and the INAF minigrant PACIFISM, respectively. CD acknowledges funding from an STFC Consolidated grant (ST/P000649/1) and a UKSA grant (ST/Y005935/1). GS acknowledges financial support under the National Recovery and Resilience Plan (NRRP), Mission 4, Component 2, Investment 1.1, Call for tender No. 104 published on 2.2.2022 by the Italian Ministry of University and Research (MUR), funded by the European Union – NextGenerationEU-Project Title 2022JC2Y93 Chemical Origins: linking the fossil composition of the Solar System with the chemistry of protoplanetary disks – CUP J53D23001600006 – Grant Assignment Decree No. 962 adopted on 30.06.2023 by the Italian Ministry of Ministry of University and Research (MUR); the project ASI-Astrobiologia 2023 MIGLIORA (“Modeling Chemical Complexity”, F83C23000800005); the INAF-GO 2024 fundings ICES, the INAF-GO 2023 fundings PROTOSKA (“Exploiting ALMA data to study planet forming disks: preparing the advent of SKA”, C13C23000770005) and the INAF Minigrant 2023 TRIESTE (“TRacing the chemIcal hEritage of our originS: from proTostars to planEts”; PI: G. Sabatini).
MP acknowledges the INAF grant 2023 MERCATOR (``MultiwavelEngth signatuRes of Cosmic rAys in sTar-fOrming Regions'')
and the INAF grant 2024 ENERGIA (``ExploriNg low-Energy cosmic Rays throuGh theoretical InvestigAtions at INAF'').
The research of MK is supported by the Carlsberg Reintegration Fellowship CF22-1014 and the DFF Sapere Aude grant 5251-00016B. SB acknowledges BASAL Centro de Astrofisica y Tecnologias Afines (CATA), project number AFB-17002. 
\\ 
In the work, we made use of {\tt astropy} \citep{astropy2018}, {\tt scipy} \citep{Virtanen2020}, {\tt numpy} \citep{Harris2020}, and {\tt PyVista} \citep{sullivan2019pyvista}.
}

\bibliographystyle{abbrvnat-maxbibnames4}
\bibliography{chapter} %

\end{document}